\documentstyle[12pt,epsfig]{article}
\def\lbldef#1#2{\expandafter\gdef\csname #1\endcsname {#2}}

\def\href#1#2{#2}  

\begin{document}
\baselineskip=15.5pt
\pagestyle{plain}
\setcounter{page}{1}
\renewcommand{\thefootnote}{\fnsymbol{footnote}}
\begin{titlepage}

\vspace*{-3cm} \today \hfill 
\raggedleft{CERN-TH/99-09\\ hep-th/9901088}

\begin{center}

\vskip 5cm

{\Large {\bf The D1-D5 Brane

System in Type I String Theory
}}

\vskip 1.5cm

{\large Yaron Oz and Donam Youm}

\vskip 0.5cm

Theory Division, CERN,
CH-1211, Geneva 23, Switzerland 

\vskip 2cm

\begin{abstract}
We construct the supergravity solution for the intersecting D1-D5 brane 
system in Type I String Theory.  The solution encodes the dependence on 
all the electric charges of the $SO(32)$ gauge group.  We discuss the 
near horizon geometry of the solution and a proposed dual $(0,4)$ 
superconformal field theory.
\end{abstract}
\end{center}
\end{titlepage}

\newpage
\def\ads{{\it AdS}}
\def\adsp{{\it AdS}$_{p+2}$}
\def\cft{{\it CFT}}

\newcommand{\beq}{\begin{equation}}
\newcommand{\eeq}{\end{equation}}
\newcommand{\ber}{\begin{eqnarray}}
\newcommand{\cN}{{\cal N}}
\newcommand{\eer}{\end{eqnarray}}
\newcommand{\cD}{{\Delta}}

\section{Introduction}

The D1-D5 brane system in Type I string theory may provide us with yet another
example of a duality between superstring theory on AdS$_3$ background and 
two dimensional SCFT. 
We will consider the system of parallel D1 and D5 branes with charges 
$Q_e$ and $Q_p$ respectively, where the D5 branes wrap a compact space $M$ 
which is $T^4$ or $K3$.  
The purpose of this letter is to construct the corresponding supergravity 
solution with non-zero $SO(32)$ gauge fields.  

The supergravity solution in Type I theory describing D1 and 
D5 branes intersecting in one spatial dimension has been constructed in 
\cite{dabhs} and partially in \cite{dabh,hull}.  However, in these solutions 
the $SO(32)$ gauge group in the open string sector is not turned on. 
Therefore, they are not complete Type I solutions, as the consistency of the 
Type I superstring theory without spacetime anomalies requires the existence 
of the $SO(32)$ gauge group.  This is what distinguishes the Type I 
supergravity solutions from the corresponding solutions in the Type IIB 
theory.  We will therefore construct the supergravity solution for the 
intersecting D-brane configuration with all the 496 electric charges of the 
$SO(32)$ gauge group turned on.  

The strategy which we will use is to first construct the supergravity 
solution for intersecting fundamental strings 
and NS5-branes with non-zero $SO(32)$ gauge field in the heterotic theory 
and then apply the $S$-duality transformation relating Type I and $SO(32)$ 
heterotic string theories to obtain the required Type I solution.  

The paper is organized as follows.  
In section 2, we briefly review the Type I - $SO(32)$ heterotic 
duality in ten dimensions. We then discuss the technique for generating 
the solution. Finally, we present the non-extreme as well the BPS heterotic
and Type I solutions. 
In section 3 we construct the near horizon geometry of the Type I solution
and discuss a proposed duality to a $(0,4)$ SCFT in two dimensions.

\section{The Supergravity Solution}

In this section, we construct the supergravity solution of Type I superstring 
theory describing intersecting D1 and D5 branes with non-zero $SO(32)$ gauge 
field.  
For this purpose, we first construct intersecting fundamental string 
and NS5-brane solution with non-zero $SO(32)$ gauge group in heterotic theory 
and then apply the $S$-duality transformation relating Type I and $SO(32)$ 
heterotic string theories to obtain this Type I theory solution.  

\subsection{Type I - $SO(32)$ heterotic duality in ten dimensions}

In this subsection, we summarize the $S$-duality transformation that 
relates the Type I superstring to the $SO(32)$ heterotic superstring in ten 
dimensions for the purpose of fixing notations.  

The bosonic part of the low energy effective action for the $SO(32)$ heterotic 
string theory is described by the massless bosonic string states, which are 
the graviton $G^{(H)}_{MN}$ ($M,N=0,1,...,9$), the dilaton $\Phi^{(H)}$ and 
the two-form field $B^{(H)}_{MN}$ in the NS sector of the closed heterotic 
string along with gauge fields $A^{(H)\,a}_{M}$ $(a=1,...,496)$ in the adjoint 
representation of the $SO(32)$ gauge group.  The action in the string frame is 
given by
\begin{eqnarray}
S^{(H)}&=&{1\over{16\pi G_{N}}}\int d^{10}x\sqrt{-G^{(H)}}
e^{-\Phi^{(H)}}\left[{\cal R}_H+G^{(H)\,MN}\partial_M\Phi^{(H)}\partial_N
\Phi^{(H)}\right.
\cr
& &\left. -{1\over{12}}H^{(H)}_{MNP}H^{(H)\,MNP}-{1\over 4}{\rm Tr}
(F^{(H)}_{MN}F^{(H)\,MN})\right],
\label{hetact}
\end{eqnarray}
where $G_{N}$ is the ten-dimensional Newton's constant, ${\cal R}_H$ is the 
Ricci scalar of the metric $G^{(H)}_{MN}$, and field strengths $H^{(H)}_{MNP}$ 
and $F^{(H)}_{MN}$ of $B^{(H)}_{MN}$ and $A^{(H)}_M$ are defined as
\begin{eqnarray}
H^{(H)}_{MNP}&=&\partial_M B^{(H)}_{NP}-{1\over 2}{\rm Tr}\left(
A^{(H)}_M F^{(H)}_{NP}-{{\sqrt{2}}\over 3}A^{(H)}_M
[A^{(H)}_N,A^{(H)}_P]\right)+{\rm cyc.\ perms.\ in\ } M,N,P,
\cr
F^{(H)}_{MN}&=&\partial_M A^{(H)}_N-\partial_N A^{(H)}_M + 
\sqrt{2}[A^{(H)}_M,A^{(H)}_N].
\label{hfsrth}
\end{eqnarray}
Here, the trace Tr is in the vector representation of $SO(32)$.  

The Type I superstring theory is defined as the orientifold projection of the 
Type IIB superstring theory.  In the massless bosonic NS sector of the closed 
string, only the graviton $G^{(I)}_{MN}$ and the dilaton $\Phi^{(I)}$ 
survive the orientifold projection.  In the bosonic RR sector, the only 
surviving massless mode is the two-form field $B^{(I)}_{MN}$.  The bosonic 
open string sector gives rise to gauge fields $A^{(I)\,a}_M$ ($a=1,...,496$) 
of the $SO(32)$ gauge group. In the string frame, these massless bosonic 
modes are described by
\begin{eqnarray}
S^{(I)}&=&{1\over{16\pi G_{N}}}\int d^{10}x\sqrt{-G^{(I)}}
\left[e^{-\Phi^{(I)}}({\cal R}_I+G^{(I)\,MN}\partial_M\Phi^{(I)}\partial_N
\Phi^{(I)})\right.
\cr
& &\left. -{1\over{12}}H^{(I)}_{MNP}H^{(I)\,MNP}-{1\over 4}e^{-{{\Phi^{(I)}}
\over 2}}{\rm Tr}(F^{(H)}_{MN}F^{(H)\,MN})\right],
\label{iact}
\end{eqnarray}
where ${\cal R}_I$ is the Ricci scalar of the metric $G^{(I)}_{MN}$, and 
field strengths $H^{(I)}_{MNP}$ and $F^{(I)}_{MN}$ of $B^{(I)}_{MN}$ and 
$A^{(I)}_M$ are defined similarly as above.

Note, these two theories have the same field contents and also the same 
supersymmetry.  In fact, the actions (\ref{hetact}) and (\ref{iact}) become 
identical, provided one relates the fields in the two actions in the 
following way \cite{witdual}:
\begin{equation}
G^{(I)}_{MN}=e^{-{{\Phi^{(I)}}\over 2}}G^{(H)}_{MN},\ \ 
\Phi^{(I)}=-\Phi^{(H)},\ \ 
B^{(I)}_{MN}=B^{(H)}_{MN},\ \ 
A^{(I)}_M=A^{(H)}_M.
\label{sdual}
\end{equation}
Since $g_s=e^{\langle\Phi\rangle}$ is defined as the string coupling constant, 
one sees that the strong coupling limit of one theory is related to the weak 
coupling limit of the other theory.  Under this $S$-duality transformations, 
the fundamental string [the solitonic NS5-brane] of the heterotic 
theory and the D-string [the D5-brane] in the Type I theory are related.

\subsection{Solution generating technique}

Fundamental strings and (solitonic) NS5-branes in heterotic string 
respectively carry electric and magnetic charges of the NS two-form field 
$B^{(H)}_{MN}$.  Additionally, we want the $p$-brane configuration to 
be charged under the $SO(32)$ gauge group.  We are therefore 
interested in the configuration which carries electric charges of 
the $SO(32)$ gauge group, i.e. the so-called {\it colored} 
non-Abelian solution
\footnote{The other class of solutions, which are regarded as genuine 
non-Abelian solutions, are {\it neutral} under the non-Abelian gauge 
group, i.e. do not carry ``global charges'' of the non-Abelian gauge group.}.  
Such solutions are constructed by embedding the $U(1)$ groups in the 
non-Abelian gauge group as the adjoint representation.  Namely, one takes 
the Ansatz for the non-Abelian $SO(32)$ gauge field $A^{(H)\,a}_M$ of the 
form $A^{(H)\,a}_M=\beta^aA_M$, where $A_M$ is an $U(1)$ gauge field 
and the parameters $\beta^a$ satisfy the constraint $\gamma_{ab}
\beta^a\beta^b=1$.  Here, $\gamma_{ab}$ is an invariant metric of the 
$SO(32)$ group. Then, the field strength $F^{(H)}_{MN}$ 
(\ref{hfsrth}) of the ``non-Abelian'' $SO(32)$ gauge group reduces 
to the field strengths $F^{(H)\,a}_{MN}=\beta^a F^0_{MN}$ of 
496 $U(1)$ gauge fields $A^{(H)\,a}_M=\beta^a A_M$.  So, the effective 
action (\ref{hetact}) becomes that with 496 Abelian $U(1)$ gauge fields.  
In particular, 496  electric charges $Q^{(H)\,a}$ of the $U(1)$ 
gauge fields form the adjoint representation of the $SO(32)$ gauge group.  
One can induce these $p$-brane charges and the electric charges of the 
$SO(32)$ gauge group on the uncharged black $p$-brane solution 
by first compactifying it down to 3 dimensions and then applying 
the appropriate boost transformations in the $U$-duality symmetry of the 
3-dimensional action.  

Since we are interested in constructing the intersecting fundamental string 
and NS5-brane solution which is localized along the overall transverse 
directions, only, we start from the following uncharged black fivebrane 
solution in $D=10$:
\begin{equation}
G^{(H)}_{MN}dx^M dx^N=-(1-{{2m}\over{r^2}})dt^2+dx^2_1+\cdots+dx^2_5+
(1-{{2m}\over{r^2}})^{-1}dr^2+r^2d\Omega^2_3,
\label{uchdbstr}
\end{equation}
where $d\Omega^{2}_{3}=d\theta^{2}+\sin^{2}\theta 
d\phi^{2}_{1}+\cos^{2}\theta d\phi^{2}_{2}$ is the infinitesimal line 
element on $S^{3}$.  The remaining fields are zero.  

One can compactify the solution (\ref{uchdbstr}) in the $t$, $x_{i}$ and 
$\phi_{2}$ directions on a torus down to three dimensions, since the 
solution is independent of these coordinates.  The Kaluza-Klein Ansatz 
of the metric for such compactification is given by
\begin{equation}
G^{(H)}_{MN}=\left(\matrix{e^{\varphi}h_{\mu\nu}+G_{mn}A^{m}_{\mu}A^{n}_{\nu}
& A^{m}_{\mu}G_{mn}\cr A^{n}_{\nu}G_{mn}& G_{mn}}\right),
\label{kkansat}
\end{equation}
where $\phi\equiv \Phi^{(H)}-{1\over 2}\ln\det G_{mn}$ is the 3-dimensional 
dilaton and the indices run as $\mu,\nu=r,\theta,\phi_1$ and 
$m,n=t,\phi_2,x_1,...,x_5$.  The ten-dimensional heterotic action 
(\ref{hetact}) compactifies to the following three-dimensional form, which 
we write down only the final form as the details on its derivation and the 
field definitions can be found elsewhere \cite{sen1,sen2,fivbh}:
\begin{equation}
{\cal L}={1\over 4}\sqrt{-h}\left[{\cal R}_{h}+{1\over 8}h^{\mu\nu}
{\rm Tr}(\partial_{\mu}{\cal M}L\partial_{\nu}{\cal M}L)\right],
\label{3dact}
\end{equation}
where ${\cal R}_h$ is the Ricci scalar of the metric $h_{\mu\nu}$.  
Note, since we keep all the 496 $U(1)$ gauge fields in the adjoint 
representation of the $SO(32)$ gauge group, the scalar moduli matrix 
${\cal M}$ is now an $O(8,504)$ matrix and $L$ is an invariant metric  
of the $O(8,504)$ group.  The action (\ref{3dact}) is invariant under 
the following $O(8,504)$ duality transformation:
\begin{equation}
h_{\mu\nu}\to h_{\mu\nu}, \ \ \ \ \ 
{\cal M}\to \Omega{\cal M}\Omega^{T},
\label{3ddual}
\end{equation}
where $\Omega\in O(8,504)$. 
The $SO(1,1)$ boost transformations in this $O(8,504)$ transformation 
group induce electric and magnetic charges of $p$-branes and 
$U(1)^{496}\subset SO(32)$ gauge group on the uncharged solution 
(\ref{uchdbstr}).  

\subsection{Non-extreme solutions}

\subsubsection{Heterotic solution}

After applying the $SO(1,1)$ boost transformations in the $O(8,504)$ 
duality group (\ref{3ddual}) with the boost parameters $\delta_e$, 
$\delta_p$ and $\delta_a$ ($a=1,...,496$) on the uncharged black 5-brane 
solution (\ref{uchdbstr}) compactified down to three dimensions, one obtains 
the following non-extreme supergravity solution describing the intersecting 
fundamental string and NS5-brane with 496 electric charges of the $SO(32)$ 
gauge group:
\begin{eqnarray}
G^{(H)}_{MN}dx^M 
dx^N&=&[1+{{m\cosh^2\delta_e(\Delta+1)-2m}
\over {r^2}}]^{-2}\left[-(1+{{2m\sinh^2\delta_e}\over{r^2}})
(1-{{2m}\over{r^2}})dt^2\right.
\cr
& &+{{2m\sinh\delta_e\cosh\delta_e(\Delta
-1)}\over {r^2}}(1-{{2m}\over{r^2}})dtdx_1
\cr
& &+\left.\left(1+{{2m(\cosh^2\delta_e\Delta -1)}
\over{r^2}}+{{m^2\cosh^2\delta_e(\Delta-1)^2}
\over{r^4}}\right)dx^2_1\right]
\cr
& &+\sum_{i=2}^5 dx^2_i +(1+{{2m\sinh^2\delta_p}\over{r^2}})
[(1-{{2m}\over{r^2}})^{-1}dr^2+r^2d\Omega^2_3],
\cr
e^{\Phi^{(H)}}&=&{{1+{{2m\sinh^2\delta_p}\over{r^2}}}\over
{1+{{m\cosh^2\delta_e(\Delta+1)-2m}\over 
{r^2}}}},
\cr
B^{(H)}_{tx_1}&=&{{{{m\sinh\delta_e\cosh\delta_e(\Delta+1)}\over{r^2}}}\over 
{1+{{m\cosh^2\delta_e(\Delta+1)-2m}
\over{r^2}}}},
\cr
B^{(H)}_{\phi_1\phi_2}&=&2m\sinh\delta_p\cosh\delta_p\sin^2\theta,
\cr
A^{(H)\,a}_t
&=&{{{{\sqrt{2}m\cosh^2\delta_e\sinh\delta_a\Delta_a}\over{r^2}}}\over
{1+{{m\cosh^2\delta_e(\Delta+1)-2m}
\over{r^2}}}},
\cr
A^{(H)\,a}_{x_1}&=&{{{{\sqrt{2}m\sinh\delta_e\cosh\delta_e\sinh\delta_a
\Delta_a}\over{r^2}}}\over 
{1+{{m\cosh^2\delta_e(\Delta+1)-2m}
\over{r^2}}}},
\label{hetsol}  
\end{eqnarray}
where 
\beq
\Delta = \prod^{496}_{i=1}\cosh\delta_i,\ \ \ \  
\Delta_a = \prod^{496}_{i=a+1}\cosh\delta_i
\eeq

The charges $Q_e$ and $Q_p$ of the fundamental string and 
the solitonic NS5-brane and the electric charges $Q_a$ of the 
$U(1)^{496}\subset SO(32)$ gauge group carried by this brane configuration 
are given by
\begin{eqnarray}
Q_e&=&m\sinh\delta_e\cosh\delta_e(\Delta+1),\ \ \ \ 
Q_p=2m\sinh\delta_p\cosh\delta_p,
\cr
Q_a&=&\sqrt{2}m\cosh^2\delta_e\sinh\delta_a\Delta_a.~~~~~~
\label{hetchg}
\end{eqnarray}

Note, in the limit where the electric charges $Q_a$ of the $SO(32)$ gauge 
group are zero ($\delta_a=0$), the solution (\ref{hetsol}) reduces to the 
non-extreme version of the intersecting fundamental string \cite{DGHR} and 
NS5-brane \cite{CHS,DR} solution.

\subsubsection{Type I solution}

The following non-extreme supergravity solution describing the intersecting 
D1 and D5 branes in the Type I string theory is obtained by applying the 
$S$-duality transformation (\ref{sdual}) on the heterotic solution 
(\ref{hetsol}):
\begin{eqnarray}
G^{(I)}_{MN}dx^{M}dx^{N}&=&\left(1+{{2m\sinh^{2}\delta_{p}}
\over{r^{2}}}\right)^{-{1\over 2}}\left[1+{{m\cosh^{2}\delta_{e}
(\Delta+1)-2m}\over{r^{2}}}\right]^{-{3\over 2}}
\cr
& &\times\left[-(1+{{2m\sinh^{2}\delta_{e}}\over{r^{2}}})
(1-{{2m}\over{r^{2}}})dt^{2}\right.
\cr
& &+{{2m\sinh\delta_{e}\cosh\delta_{e}(\Delta
-1)}\over{r^{2}}}(1-{{2m}\over{r^{2}}})dtdx_{1}
\cr
& &+\left.\left(1+{{2m(\cosh^{2}\delta_{e}\Delta
-1)}\over{r^{2}}}+{{m^{2}\cosh^{2}\delta_{e}
(\Delta-1)^{2}}\over {r^{4}}}\right)
dx^{2}_{1}\right]
\cr
& &+\left[{{1+{{m\cosh^{2}\delta_{e}(\Delta
+1)-2m}\over{r^{2}}}}\over{1+{{2m\sinh^{2}\delta_{p}}
\over{r^{2}}}}}\right]^{1\over 2}(\sum_{i=2}^5 dx^{2}_{i})
\cr
& &+\left(1+{{2m\sinh^{2}\delta_{p}}\over{r^{2}}}\right)^{1\over2}
\left[1+{{m\cosh^{2}\delta_{e}(\Delta+1)-2m}\over
{r^{2}}}\right]^{1\over 2}
({{dr^{2}}\over{1-{{2m}\over{r^{2}}}}}+r^2d\Omega^{2}_{3}),
\cr
e^{\Phi^{(I)}}&=&{{1+{{m\cosh^{2}\delta_{e}(\Delta
+1)-2m}\over{r^{2}}}}\over{1+{{2m\sinh^{2}\delta_{p}}
\over{r^{2}}}}},
\label{typisol}
\end{eqnarray}
where the remaining fields have the same forms as the heterotic 
solutions (\ref{hetsol}).  

The D1 brane charge $Q_e$, the D5 brane charge $Q_p$ and electric charges 
$Q_a$ of the $U(1)^{496}\subset SO(32)$ gauge group are given as in 
(\ref{hetchg}). 

In the limit that the electric charges $Q_a$ of the $SO(32)$ gauge group 
are zero ($\delta_a=0$), the solution (\ref{typisol}) reduces to the 
non-extreme generalizaton of the intersecting D1 and D5 brane solution 
constructed in \cite{dabhs}. 

\subsection{The BPS solutions}

The BPS limit of the above solutions (\ref{hetsol}) and (\ref{typisol}) 
is defined as the limit in which the non-extremality parameter $m$ goes 
to zero while keeping $2m\sinh\delta_e\cosh\delta_e\equiv Q$ and 
$2m\sinh\delta_p\cosh\delta_p=Q_p$ as finite non-zero constants.  
In the following, we write down the $p$-brane solutions (\ref{hetsol}) and 
(\ref{typisol}) in the BPS limits.  

\subsubsection{Heterotic solution}

The BPS intersecting fundamental string and NS5-brane solution with 
the electric charges of the $SO(32)$ gauge group is given by:
\begin{eqnarray}
G^{(H)}_{MN}dx^Mdx^N&=&\left[1+{{Q(\Delta+1)}
\over{2r^2}}\right]^{-2}\left[-(1+{Q\over{r^2}})dt^2\right.
\cr
& &+\left.{{Q(\Delta-1)}\over{r^2}}dtdx_1
+(1+{{Q\Delta}\over{r^2}})dx^2_1\right]
\cr
& &+\sum_{i=2}^5 dx^2_i+(1+{{Q_p}\over{r^2}})(dr^2+r^2d\Omega^2_3),
\cr
e^{\Phi^{(H)}}&=&(1+{{Q_p}\over{r^2}})(1+{{Q(\Delta+1)}
\over {2r^2}})^{-1},
\cr
B^{(H)}_{tx_1}&=&{{Q(\Delta+1)}\over{2r^2}}
(1+{{Q(\Delta+1)}\over{2r^2}})^{-1},\ \ \ 
B^{(H)}_{\phi_1\phi_2}=Q_p\sin^2\theta,
\cr
A^{(H)\,a}_t&=&{{Q_a}\over{r^2}}(1+{{Q(\Delta+1)}
\over{2r^2}})^{-1}=A^{(H)\,a}_{x_1}.
\label{bpshetsol}
\end{eqnarray}
Note, the charge $Q_e$ of the fundamental string is defined as 
$Q_e=Q(\Delta +1)$.

\subsubsection{Type I solution}

The BPS intersecting D1 and D5 brane with electric charges of the 
$SO(32)$ gauge group in the open string sector is as follows:
\begin{eqnarray}
G^{(I)}_{MN}dx^Mdx^N&=&(1+{{Q_p}\over{r^2}})^{-{1\over 2}}
(1+{{Q(\Delta+1)}\over{2r^2}})^{-{3\over 2}}
\left[-(1+{Q\over{r^2}})dt^2\right.
\cr
& &+\left.{{Q(\Delta-1)}\over{r^2}}dtdx_1
+(1+{{Q\Delta}\over{r^2}})dx^2_1\right]
\cr
& &+(1+{{Q(\Delta+1)}\over{2r^2}})^{1\over 2}
(1+{{Q_p}\over{r^2}})^{-{1\over 2}} \sum_{i=2}^5 dx^2_i
\cr
& &+(1+{{Q_p}\over{r^2}})^{1\over 2}(1+{{Q(\Delta+1)}
\over {2r^2}})^{1\over 2}(dr^2+r^2d\Omega^2_3),
\cr
e^{\Phi^{(I)}}&=&(1+{{Q(\Delta+1)}\over{2r^2}})
(1+{{Q_p}\over{r^2}})^{-1},
\label{bpstypisol}
\end{eqnarray}
where the D1 brane charge $Q_e$ is given by $Q_e=Q(\Delta +1)$.

\section{The Near-Horizon Geometry and $(0,4)$ SCFT}

The decoupling limit of the worldvolume theory of the intersecting $D$-brane 
configuration corresponds to the near-horizon limit of the corresponding 
supergravity solution.  The near-horizon geometry of the supergravity 
solution (\ref{bpstypisol}) is obtained by keeping only the $1/r^2$ 
terms in the harmonic functions in the solution.  The resulting metric 
has the following form:
\begin{eqnarray}
G^{(I)}_{MN}dx^Mdx^N&=&\left({2\over{\Delta+1}}
\right)^{3\over 2}{{r^2}\over\sqrt{QQ_p}}\left[-dt^2+(\Delta-1)dtdx_1
+\Delta dx^2_1\right]
\cr
& &+\sqrt{{\Delta+1}\over 2}\sqrt{Q\over{Q_p}}
(dx^2_2+\cdots+dx^2_7)
\cr
& &+\sqrt{{\Delta+1}\over 2}\sqrt{QQ_p}
\left({{dr^2}\over{r^2}}+d\Omega^2_3\right).
\label{nearhor1}
\end{eqnarray}
If one redefines the coordinates $t$ and $x_1$ in the following way:
\begin{eqnarray}
t&\to&t^{\prime}=\sqrt{2\over{\Delta+1}}t,
\ \ \ \ \ \ \ \ \ 
x_1\to x^{\prime}_1=\sqrt{2\over{\Delta+1}}x_1,
\cr
t^{\prime}&\to&t^{\prime\prime}=t^{\prime}-
{{\Delta-1}\over 2}x^{\prime}_1,
\ \ \ 
x^{\prime}_1\to x^{\prime\prime}_1={{\Delta+1}\over 2}
x^{\prime}_1,
\label{coordtr}
\end{eqnarray}
then the metric (\ref{nearhor1}) takes the recognizable form 
(in the following double primes are suppressed):
\begin{equation}
G^{(I)}_{MN}dx^Mdx^N={{r^2}\over{\sqrt{Q_eQ_p}}}(-dt^2+dx^2_1)+
{{\sqrt{Q_eQ_p}}\over{r^2}}dr^2+\sqrt{{Q_e}\over{Q_p}} \sum_{i=2}^5 dx^2_i
+\sqrt{Q_eQ_p}d\Omega^2_3.
\label{nearhor2}
\end{equation}
This corresponds to the metric describing $AdS_3\times M \times S^3$ with  
the radii of $AdS_3$ and $S^3$ being $R^2_{AdS}=R^2_{S^3}=\sqrt{Q_eQ_p}$ and 
the volume of $M$ being $v_{M}=Q_e/Q_p$.  

The D1-D5 brane configuration in Type I string theory is expected
to encode the information
on the moduli space of instantons on $M$. 
The worldvolume theory of $k$ flat D5-branes in Type I theory
has as its Higgs branch the moduli space of $SO(32)$ $k$-instantons on
$R^4$ \cite{witten1}.  
Using a D1 brane probe extending parallel to the D5-branes, one gets
the ADHM data encoded in the Yukawa couplings of the $(0,4)$ sigma model of 
D1 brane worldvolume theory \cite{douglas, witten2}.
More precisely, the condition for $(0,4)$ supersymmetry of the sigma model 
requires that the couplings satisfy the ADHM equations.
For every instanton, one gets a $(0,4)$ sigma model that flows in the infrared
to a solution of string theory for the corresponding instanton.
This generalizes to other non-compact manifolds such as ALE spaces where a 
similar ADHM construction exists.

When the D5 branes wrap a compact manifold $M$, there should presumably be 
a similar relation between the $(0,4)$ theory on the D1 brane worldvolume 
to the instanton moduli space.  This is not yet known.
In the spirit of other examples of the AdS/SCFT correspondence 
\cite{maldacena,GKP,witten} one is led to conjecture that the $(0,4)$ SCFT 
in the infrared limit of the D1 brane worldvolume theory is
dual to Type I string theory on the $AdS_3\times M \times S^3$ background 
(\ref{nearhor2})\footnote{See also \cite{Cliff,Jose}.}.  
The supergroup $SU(1,1/2) \times SL(2,R) \times SU(2)$ of 
the Type I compactification is mapped under the conjectured duality to the 
symmetry supergroup of the $(0,4)$ SCFT.

For charges $Q_e,Q_p >> 1$, one can use the supergravity approximation.
The supergravity theory can be reduced first on $M$ from ten to six 
dimensions.  One can then study the spectrum of the Kaluza-Klein 
excitations of $N=1$ supergravity on $AdS_3 \times S^3$ as in \cite{jdb}.
It will be interesting to construct the string theory picture 
as done for the D1-D5 system in Type IIB string theory \cite{gks}.
 
\section*{Acknowledgments}

We would like to thank N. Nekrasov and S. Shatashvili
for discussions.

\end{document}